\documentclass[%
reprint,
 amsmath,amssymb,
 aps,
prb,
twocolumn,
superscriptaddress,
longbibliography
]{revtex4-2}

\usepackage{graphicx}
\usepackage{dcolumn}
\usepackage{bm}
\usepackage{txfontsb}
\usepackage[dvipsnames]{xcolor}
\usepackage[T1]{fontenc}
\newcommand{\AffA}{Department of Applied Physics, The University of Tokyo, Tokyo 113-8656, Japan}

\begin{document}

\preprint{AAPM/123-QED}

\title[Sample title]{Effect of interaction range on the phase diagram of magnetic hedgehog lattices}

\author{Midori Yamada}
 \email{midori-yamada@g.ecc.u-tokyo.ac.jp}
\author{Kotaro Shimizu}

\affiliation{\AffA}

\author{Shun Okumura}
\affiliation{Quantum-Phase Electronics Center (QPEC), The University of Tokyo, Tokyo 113-8656, Japan} 
\affiliation{RIKEN Center for Emergent Matter Science (CEMS), Wako 351-0198, Japan}
\affiliation{International Institute for Sustainability with Knotted Chiral Meta Matter (WPI-SKCM$^2$),
Hiroshima University, Hiroshima 739-8531, Japan}
\author{Yasuyuki Kato}
\affiliation{
Department of Applied Physics, University of Fukui, Fukui 910-8507, Japan
}%
\author{Yukitoshi Motome}
\affiliation{\AffA}

\date{\today}

\begin{abstract}

Topological spin textures have attracted considerable attention through their emergent electromagnetic properties, including the topological Hall effect and the topological Nernst effect. Among them, magnetic hedgehog lattices (HLs) are three-dimensional topological spin textures that host emergent magnetic monopoles and antimonopoles. HLs have been identified in metallic compounds, such as $\mathrm{MnSi}_{1-x}\mathrm{Ge}_x$ and $\mathrm{SrFeO_3}$. Various theoretical models have been proposed, with either short-range or long-range interactions, which are typically regarded as effective descriptions of insulating and metallic systems, respectively.
However, the role of the interaction range in stabilizing the HLs remains only partially understood.
To address this gap, we investigate the stability of HLs by systematically varying the range and spatial decay of the exchange interactions, thereby interpolating between interaction profiles commonly associated with localized-spin insulating models and itinerant-electron metallic models. 
Based on extensive Monte Carlo simulations, we find that HLs are stabilized over a broad interaction range. Intriguingly, among the interaction profiles examined, intermediate-range interactions are particularly favorable for stabilizing HLs. 
They give rise to two distinct types of $3Q$-HLs that differ in the number of monopole-antimonopole pairs per magnetic unit cell, and their relative stability depends sensitively on the interaction range. 
Our results provide a unified framework for tuning phase diagrams from microscopic interaction models, serving as a guideline for interpreting experiments in candidate materials.

\end{abstract}

\maketitle

\section{Introduction}

Topologically nontrivial spin textures can generate emergent electromagnetic fields through the real-space Berry phase, arising from their noncoplanar spin configurations \cite{nagaosa2013topological-d14}. These emergent fields can give rise to unconventional transport phenomena, such as the topological Hall effect \cite{PhysRevB.62.R6065, doi:10.1126/science.1058161, AHE}.
A paradigmatic example is the magnetic skyrmion, a two-dimensional (2D) swirling spin texture \cite{muhlbauer2009skyrmion,PhysRevLett.102.186602, yu2010real,nagaosa2013topological-d14}. 
Meanwhile, three-dimensional (3D) topological spin textures, such as magnetic hedgehogs \cite{kanazawa2011large, tanigaki2015real, fujishiro_hl, ishiwata_hl} and hopfions \cite{kent2021creation, yu2023realization, zheng2023hopfion}, have attracted increased attention. 
Their additional spatial degrees of freedom and topological structures  \cite{doring1968point, L_Niemi_1997} may support dynamical and transport phenomena beyond those of 2D skyrmions \cite{kanazawa2011large, PhysRevB.88.064409, fujishiro2018large,LBo_spinexc_hopfion, khodzhaev2022hopfion, yu2023realization, Xu_Deenen_Grundler_2026, v929-88l7, ngvg-h27j}. They have therefore been proposed as promising building blocks for multidimensional information encoding in future spintronic devices \cite{Ruiz-Gomez_2025, Farinha_Yang_Yoon_Pal_Parkin_2025, skroadmap2026}.

\begin{figure}
\includegraphics[width=\columnwidth]{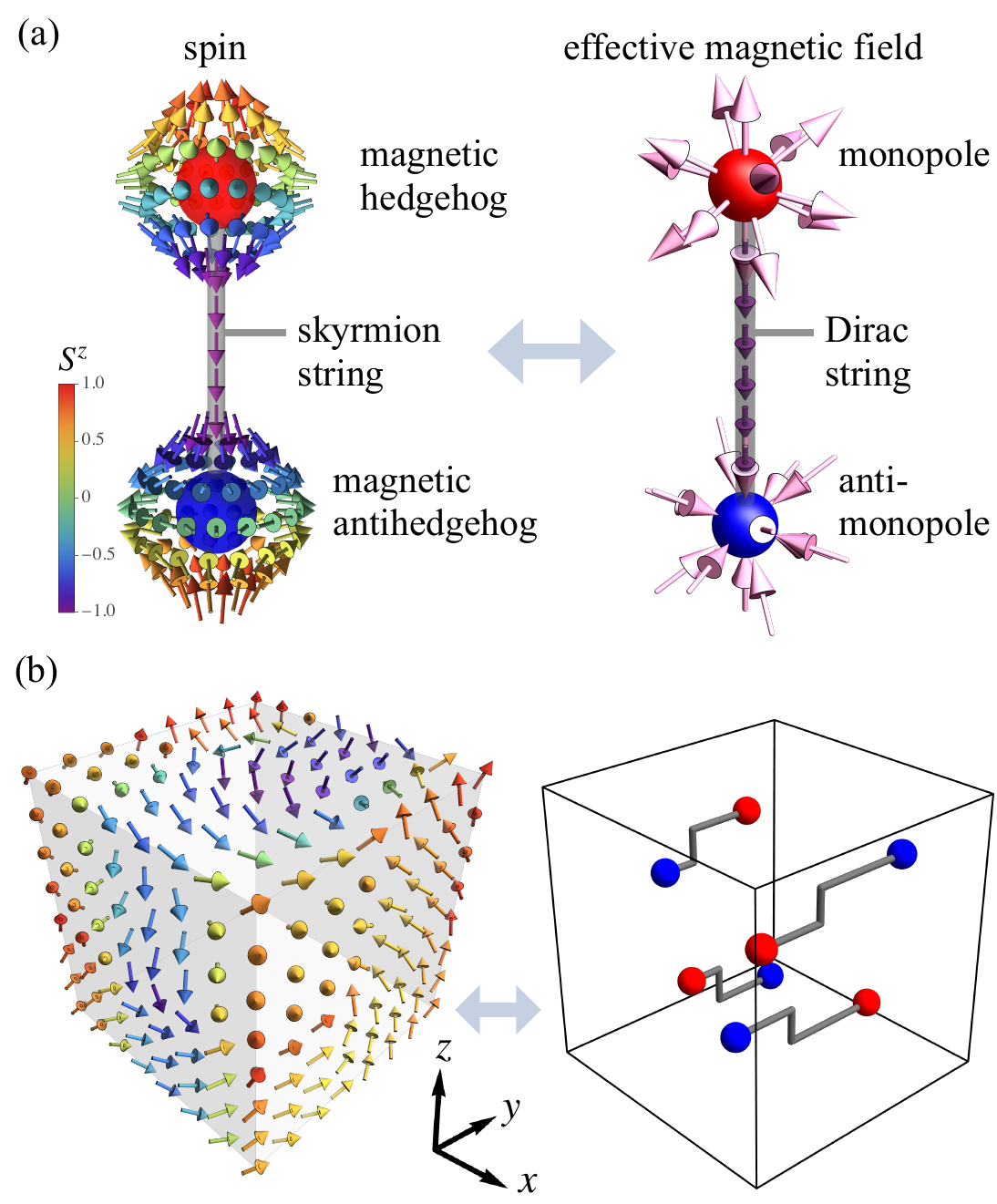}
\caption{\label{fig:HL} 
(a) Schematic illustrations of a magnetic toron in real space, consisting of a magnetic hedgehog (red sphere) and a magnetic antihedgehog (blue sphere) connected by a skyrmion string (gray tube) (left). The arrows represent the spins, and their color indicates the $z$ component.
The right panel represents the corresponding effective magnetic field, showing that a toron can be regarded as a monopole–antimonopole pair (red and blue spheres) connected by a Dirac string (gray tube).
(b) Real-space spin configuration of a 3D topological spin structure known as a HL (left), and the corresponding real-space distribution of magnetic torons within a magnetic unit cell (right).
}
\end{figure}

In a continuum description, magnetic hedgehogs and antihedgehogs have singularities at their cores, where the spin length vanishes. They can therefore act as sources and sinks of the emergent magnetic field, respectively, making them analogous to emergent magnetic monopoles and antimonopoles \cite{doring1968point, mildemonopole}. 
The associated monopole charge is a topological invariant determined by the flux of the emergent magnetic field through a surface enclosing the core singularity \cite{Volovik_1987, N_Y_X_A_G_F_Y_N_Y_2016}.
In spin systems, a magnetic hedgehog and antihedgehog often appear as a pair connected by a skyrmion string, forming a so-called magnetic toron \cite{PhysRevB.98.054404}. This object can be regarded as a monopole–antimonopole pair connected by a Dirac string, as illustrated in Fig.~\ref{fig:HL}(a).

Magnetic hedgehogs have been identified in metallic magnets, including the $B20$ compounds $\mathrm{MnGe}$ \cite{kanazawa2011large, tanigaki2015real,kanazawa_hl, kanazawa2017noncentrosymmetric} and $\mathrm{MnSi}_{1-x}\mathrm{Ge}_x$ \cite{ fujishiro_hl}, as well as in the perovskite compound $\mathrm{SrFeO_3}$ \cite{PhysRevB.84.054427, ishiwata_hl}.
In these materials, hedgehogs and antihedgehogs are periodically arranged, forming magnetic hedgehog lattices (HLs).
One representative example is the $3Q$-HL, which is described by the superposition of three orthogonally arranged helical spin waves \cite{PhysRevB.83.184406,kanazawa_hl,kanazawa2011large, tanigaki2015real}. Figure~\ref{fig:HL}(b) shows the spin texture of a $3Q$-HL together with a schematic real-space arrangement of magnetic torons within one magnetic unit cell. 
Another example is the $4Q$-HL, which is characterized by the superposition of four tetrahedrally aligned spin helices \cite{PhysRevB.83.184406,fujishiro_hl, PhysRevLett.132.226705, ishiwata_hl, yang_hlsimulation}. 
In real compounds, the modulation periods of these $3Q$- and $4Q$-HLs are often only a few nanometers.
These short modulation lengths are understood to create a large emergent magnetic field, which can induce giant topological Hall and thermoelectric responses \cite{kanazawa2011large, PhysRevB.88.064409, fujishiro2018large, fujishiro_hl}.

Various theoretical models have been proposed to characterize the stabilization mechanism of HLs.
One example is provided by localized-spin models, where the dominant interactions are typically assumed to be short-ranged. Their interactions can originate from direct exchange or superexchange processes between neighboring magnetic ions, often mediated by ligand orbitals \cite{anderson1950antiferromagnetism,goodenough1955, KANAMORI195987}. 
The interactions can also originate from spin-orbit coupling, giving rise to anisotropic exchange terms, including the antisymmetric Dzyaloshinskii-Moriya (DM) interaction, when inversion symmetry is absent at the relevant bond center \cite{DZYALOSHINSKY, PhysRev.120.91}.
Within this framework, HLs have been stabilized through short-range ferromagnetic (FM) and DM exchange interactions \cite{yang_hlsimulation, S_J_M_J_O_G_S_Y_S_2020}, competing antiferromagnetic interactions on geometrically frustrated lattices \cite{PhysRevB.103.014406}, or chiral interactions \cite{S_J_M_J_O_G_S_Y_S_2020, maysprb2026}. 

A complementary approach employs spin-charge-coupled itinerant-electron models, such as the Kondo-lattice model.
In the weak-coupling regime, a perturbative expansion in the spin-charge coupling yields, at second-order, long-range, oscillatory Ruderman-Kittel-Kasuya-Yosida (RKKY) interactions between localized moments \cite{PhysRev.96.99, PTP.16.45, PhysRev.106.893}.
This interaction favors modulations characteristic of the electronic structure, e.g., those associated with enhanced electronic susceptibility due to Fermi-surface nesting.
Effective models that retain only these dominant modes correspond to long-range interactions in real space \cite{PhysRevB.95.224424, PhysRevLett.121.137202}. 
The fourth-order contributions generate higher-order spin interactions, often represented by biquadratic interactions, thereby favoring multiple-$Q$ magnetic order \cite{Hayami_2021, PhysRevLett.108.096401, PhysRevB.95.224424}.
Such bilinear-biquadratic models, together with DM interaction, have been shown to stabilize $3Q$-HLs  \cite{PhysRevB.101.144416}, and even $4Q$-HLs in the centrosymmetric case \cite{10.7566/JPSJ.91.093702}.
$3Q$-HLs can also be stabilized within effective bilinear spin models containing momentum-dependent anisotropic exchange interactions \cite{PhysRevB.105.174413}.

Short-range interactions are commonly used to describe localized-spin insulating systems, whereas long-range interaction models based on RKKY-type interactions are typically employed for metallic systems.
To date, all the candidate materials discussed above in which HLs have been identified are metallic. 
Nevertheless, HLs may emerge in insulating compounds with short-range interactions, as noted above.
Moreover, exchange interactions in real materials exhibit material-dependent ranges and spatial dependences.
Their effective strengths and spatial profiles may be renormalized by electronic correlations, zero-point spin fluctuations, and finite-temperature magnetic and lattice fluctuations \cite{1973639, Liechtenstein_1984, PhysRevLett.111.127204, PhysRevResearch.2.043357, PhysRevB.91.125133}. 

In this paper, we investigate how the spatial range of exchange interactions affect the stability of HLs.
To this end, we compare a series of interaction profiles, ranging from rapidly decaying, short-range couplings to slowly decaying, long-range couplings.
Monte Carlo (MC) simulations show that $3Q$-HLs can be stabilized over a broad range of interaction profiles.
Notably, among the cases examined, the intermediate-range interactions stabilize two distinct types of $3Q$-HLs under an applied magnetic field. These HL phases are distinguished by the different number of monopole-antimonopole pairs, i.e., the toron number per magnetic unit cell, and their relative stability depends on the interaction range.
Our results establish a systematic connection between microscopic interaction profiles and the stability of HLs, providing guidance for further exploration of candidate materials.

The rest of this paper is structured as follows.  
Section~\ref{subsec:model} introduces the Hamiltonian of the effective bilinear spin model for infinite- and finite-range interactions, and Sec.~\ref{subsec:methods} describes the numerical method employed in this study. In Sec.~\ref{sec:results}, we present the magnetic field-temperature phase diagrams for different interaction ranges and characterize the obtained phases. Sections~\ref{subsec:short-range}, \ref{subsec:long-range}, and \ref{subsec:intermediate-range} focus on the short-, infinite-, and intermediate-range interaction cases, respectively. In Sec.~\ref{sec:discussion}, we discuss the stabilization mechanisms and general trends of HLs over the different interaction ranges.
Finally, Sec.~\ref{sec:conclusion} is devoted to the conclusion of this study.

\section{Model and Methods}
\label{sec:modelandmethods}

\subsection{Effective spin model}
\label{subsec:model}

 We introduce an effective bilinear spin model for cubic chiral magnets, following previous studies~\cite{PhysRevB.104.224405, PhysRevB.105.174413, shimizu2025currentinduced-01a}. 
Our starting point is the metallic limit, in which itinerant electrons are coupled to localized magnetic moments. 
Within a weak-coupling expansion, the second-order contribution provides an effective spin model with long-range and oscillatory interactions between the localized moments through the RKKY mechanism \cite{PhysRev.96.99, PTP.16.45,PhysRev.106.893}. 
In Kondo-lattice-type models, the corresponding momentum-space interactions can be enhanced at characteristic wave vectors, such as those associated with Fermi-surface nesting vectors \cite{PhysRevB.95.224424,PhysRevLett.121.137202}. 
Retaining only the dominant interaction components at these wave vectors yields an effective model with infinite-range interactions, whose Hamiltonian is written in momentum space as
\begin{equation} 
     \mathcal{H}_0 = \sum_{\eta} \Biggl[\sum_{\alpha,\beta}-J^{\alpha\beta}_{\mathbf{Q}_\eta} S^\alpha_{\mathbf{Q}_\eta}  S^\beta_{-\mathbf{Q}_\eta} 
     -i\mathbf{D}_\eta \cdot (\mathbf{S}_{\mathbf{Q}_\eta} \times \mathbf{S}_{-\mathbf{Q}_\eta} ) \Biggr] - \sum_i HS^z_{i}.
     \label{eq:infrange_hami}
\end{equation}
Here, $\alpha, \beta = x,y,z$, and the sum over $\eta$ is taken over the ordering wave vectors $\mathbf{Q}_{\eta}$. 
The first term represents the symmetric exchange interaction, while the second term describes the antisymmetric interaction of DM type. The last term denotes the Zeeman coupling to the external magnetic field $\mathbf{H} = (0,0,H)$. 
In Eq.~\eqref{eq:infrange_hami}, $\mathbf{S_q} = \frac{1}{\sqrt{N}} \sum_i\mathbf{S}_ie^{-i\mathbf{q}\cdot\mathbf{r}_i}$ is the Fourier transform of the localized moments $\mathbf{S}_i = (S^x_{\mathbf{r}_i}, S^y_{\mathbf{r}_i},S^z_{\mathbf{r}_i})$ at lattice site $i$ located at the real-space position $\mathbf{r}_i$, where $N$ is the total number of spins. 
We consider a simple cubic lattice with lattice constant equal to unity.
We treat the localized moments as classical unit vectors $|\mathbf{S}_i| = 1$.
In this study, to stabilize $3Q$-HLs, we incorporate three ordering vectors $\mathbf{Q}_1 = (Q,0,0), \mathbf{Q}_2 = (0,Q,0)$, and $ \mathbf{Q}_3 = (0,0,Q)$, and set $Q=\pi/4$ with the magnetic period of $8$ lattice sites.

For simplicity, we assume the symmetric exchange interactions to contain only the diagonal components $J^{\alpha\alpha}_{\mathbf{Q}_\eta}$. We also take each DM vector to be parallel to the corresponding wave vector, $\mathbf{D}_{\eta}\parallel \mathbf{Q}_{\eta}$, which favors proper-screw-type spin modulations. 
In addition, we introduce an exchange anisotropy $\Delta$ in the symmetric interaction, while preserving the $C_3$  rotational symmetry about the $[111]$ direction, which relates $\mathbf{Q}_1$, $\mathbf{Q}_2$ and $\mathbf{Q}_3$. 
This anisotropy deforms the spin helix from circular to elliptical. 
With these assumptions, Eq.~\eqref{eq:infrange_hami} reduces to
\begin{equation}
     \mathcal{H}_0 = \sum_{\eta}\sum_{\alpha,\beta} \mathrm{B}^{\alpha\beta}_{\mathbf{Q}_\eta} S^\alpha_{-\mathbf{Q}_\eta}  S^\beta_{\mathbf{Q}_\eta} - \sum_i HS^z_{i},
     \label{eq:H_infrange}
\end{equation}
 with 
\begin{equation}
   \mathrm{B}_{\mathbf{Q}_1} = \begin{pmatrix}
    -J(1-\Delta) & 0 & 0\\
    0 & -J(1+2\Delta) & iD \\
    0 & -iD & -J(1-\Delta)
    
\end{pmatrix},
\end{equation}
\begin{equation}
   \mathrm{B}_{\mathbf{Q}_2} = \begin{pmatrix}
    -J(1-\Delta) & 0 & -iD \\
    0 & -J(1-\Delta) & 0 \\
    iD & 0 & -J(1+2\Delta)
\end{pmatrix},
\end{equation}
\begin{equation}
   \mathrm{B}_{\mathbf{Q}_3} = \begin{pmatrix}
    -J(1+2\Delta) & iD & 0\\
    -iD & -J(1-\Delta) & 0 \\
    0 & 0 & -J(1-\Delta)
\end{pmatrix}.
\end{equation} 
These matrices are related by cyclic permutations of the $x$, $y$, and $z$ axes.
We set $J = 1$ as the energy unit, and take $\Delta = 0.3$ and $D = 0.3$, a parameter set known to stabilize a $3Q$-HL.

\begin{figure}
\includegraphics[width=0.9\columnwidth]{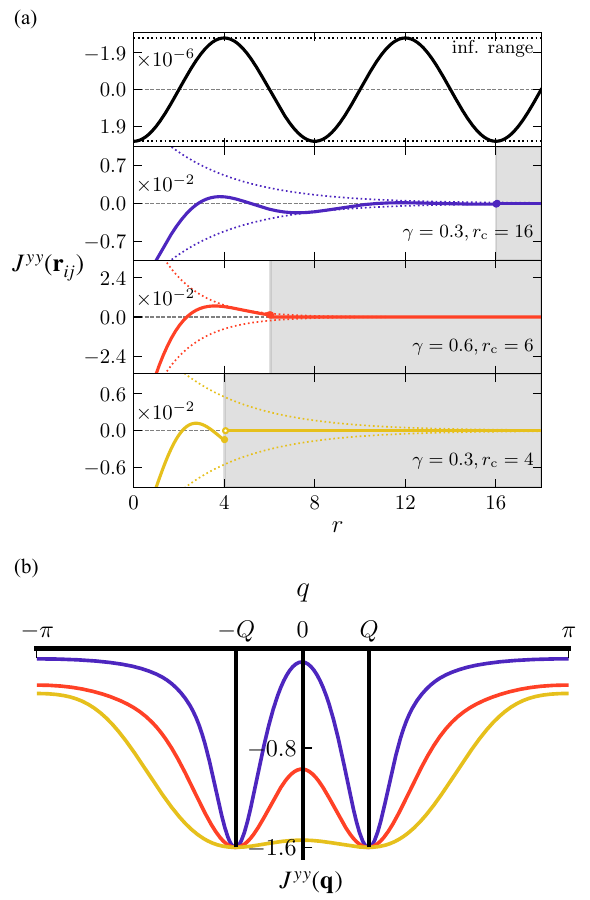}
\caption{\label{fig:Jr_Jq} Exchange interaction $J^{yy}$ in (a) real space along the $x$ axis with $\mathbf{r}_{ij} = (x,0,0)$ and (b) momentum space along the $q_x$ axis with $\mathbf{q}= (q,0,0)$, for different decay strengths and cutoff lengths: infinite interaction range (black), $\gamma = 0.3$ and $r_\mathrm{c} =16$ (blue), $\gamma = 0.6$ and $r_\mathrm{c} =6$ (red), and $\gamma = 0.3$ and $r_\mathrm{c} = 4$ (yellow). The gray region in (a) marks distances beyond the cutoff length, above which the interactions are set to zero.}
\end{figure}

In addition to the infinite-range interaction model in Eq.~\eqref{eq:H_infrange}, we consider a finite-range interaction model by introducing spatial decay into the interactions \cite{PhysRevB.104.224405, shimizu2025currentinduced-01a}. 
The decay is controlled by the parameter $\gamma$, which acts as an exponential damping factor, as illustrated in Fig.~\ref{fig:Jr_Jq}. 
For the finite-range interaction model, we employ the real-space Hamiltonian given by
\begin{equation}
    \mathcal{H} = \sum_{i,j, g(\mathbf{r}_{ij})\leq r_\mathrm{c}}\sum_{\alpha\beta} S^\alpha_i J^{\alpha\beta}_{ij} S^\beta_j - \sum_i HS^z_{i},
    \label{eq:H}
\end{equation}
with
$g(\mathbf{r}_{ij})= |r_{ij}^x| + |r_{ij}^y| + |r_{ij}^z|$.
In addition to introducing an exponential decay, we impose a cutoff at $g(\mathbf{r}_{ij})\leq r_{\mathrm{c}}$ for computational simplicity, neglecting all interactions beyond this range. 
The exchange interactions are defined as 
\begin{align}
J_{ij}=J(\mathbf{r}_{ij})
={}&
\frac{2e^{-\gamma g(\mathbf{r}_{ij})}}{L_\gamma(0)^3}
\sum_{\eta=1}^{3}
\Biggl[
    \operatorname{ReB}_{\mathbf{Q}_\eta}
    \left(
        \cos(\mathbf{Q}_{\eta}\cdot\mathbf{r}_{ij})
        - \bar{L}_\gamma(\mathbf{Q}_{\eta})
    \right)
\notag\\
&\qquad\qquad\quad
    - \operatorname{ImB}_{\mathbf{Q}_\eta}
      \sin(\mathbf{Q}_{\eta}\cdot\mathbf{r}_{ij})
\Biggr] ,
\label{eq:Jij_finite}
\end{align}
where 
\begin{equation}
    \bar{L}_\gamma(\mathbf{q}) = \frac{L_\gamma(q_x)L_\gamma(q_y)L_\gamma(q_z)}{L_\gamma(0)^3},
    \label{eq:lbar}
\end{equation}
with
\begin{equation}
    L_\gamma(q) = \frac{\mathrm{sinh}\gamma}{\mathrm{cosh}\gamma -\mathrm{cos}q}.
\end{equation}

Truncating the real-space interaction can shift the peak positions of the momentum-space interaction $J(\mathbf{q})$ obtained by Fourier transforming Eq.~\eqref{eq:Jij_finite}.
To recover the dominant contributions to be centered at the target ordering wave vectors $\mathbf{q}=\pm\mathbf{Q}_{\eta}$, we replace $Q$ in Eq.~\eqref{eq:Jij_finite} with an adjusted wave number $Q^{*}$, following Ref.~\cite{PhysRevB.104.224405}. 
We determine $Q^{*}$ numerically.
For each trial value of $Q^{*}$, we Fourier transform Eq.~\eqref{eq:Jij_finite} to obtain $J_{\gamma,r_{\mathrm{c}},Q^{*}}(\mathbf{q})$ and locate its peak position. 
We then adjust $Q^{*}$ using the bisection method until the peak positions coincide with the target ordering wave vector $\mathbf{Q}_{\eta}$.
Finally, we inverse Fourier transform the optimized $J_{\gamma,r_{\mathrm{c}},Q^{*}}(\mathbf{q})$ to obtain the real-space interaction $J(\mathbf{r}_{ij})$ used in the following calculations. 
The subtraction term $\bar{L}_\gamma(\mathbf{Q}_{\eta})$ in Eq.~\eqref{eq:Jij_finite} reduces contributions from $\mathbf{q}=0$ \cite{shimizu2025currentinduced-01a}.

To examine the effect of the interaction range, we consider the parameter sets ($\gamma, r_\mathrm{c}$) = ($0.3,4$), ($0.6,6$), and ($0.3,16$), in addition to the infinite-range case [Eq.~\eqref{eq:H_infrange}].
To quantify the spatial extent of the interactions, we define the effective interaction range as
\begin{equation}
    r_{\mathrm{eff}}(\gamma, r_{\mathrm{c}}) = \frac{\sum_{0<g(\mathbf{r})\leq r_\mathrm{c}}g(\mathbf r)\|J(\mathbf r)\|_{\mathrm F}}{\sum_{0<g(\mathbf{r})\leq r_\mathrm{c}}\|J(\mathbf r)\|_{\mathrm F}},
    \label{eq:reff}
\end{equation}
where $\|\dots\|_{\mathrm F}$ denotes the Frobenius norm and the sums include only interaction terms within the cutoff length $r_\mathrm c$.
The resulting effective ranges are $r_{\mathrm{eff}}=2.47$, $3.10$, and $8.22$ for $(\gamma,r_\mathrm{c})=(0.3,4)$, $(0.6,6)$, and $(0.3,16)$, respectively. Thus, the spatial extent of the finite-range interactions increases in the order $r_{\mathrm{eff}}(0.3,4)<r_{\mathrm{eff}}(0.6,6
)<r_{\mathrm{eff}}(0.3,16)$.
These parameter sets are also chosen such that the $3Q$-HL phase is stabilized at zero magnetic field. 
Figure~\ref{fig:Jr_Jq} shows the resulting real-space interaction component $J^{yy}({\mathbf{r}}_{ij})$ along the $x$ direction and its momentum-space counterpart $J^{yy}({\mathbf{q}})$ along the $q_x$ direction. 
More details of the interaction profiles are provided in Appendix~\ref{app:interaction}.

Hereafter, we refer to the model with $(\gamma, r_\mathrm{c}) = (0.3,4)$ as the short-range interaction model and those with $(\gamma, r_\mathrm{c}) = (0.6,6)$ and $(0.3,16)$ as the intermediate-range interaction models. 

\subsection{Methods}
\label{subsec:methods}

We examine the phase diagrams of the models defined by Eqs.~\eqref{eq:H_infrange} and \eqref{eq:H} using parallel-tempering MC simulations based on the standard Metropolis algorithm.
Parallel tempering mitigates trapping in local minima by allowing replica exchanges between different temperatures \cite{doi:10.1143/JPSJ.65.1604}.

For each parameter set, we use $64$ replicas at temperatures geometrically distributed between $T_{\mathrm{min}} = 0.2$ and $T_{\mathrm{max}} = 1.6$.
We first perform $10^6$ MC sweeps for equilibration, attempting replica exchanges every $10^3$ sweeps. We then measure the observables over an additional $5\times 10^5$ MC sweeps. The measurement is divided into $N_{\mathrm b}=1000$ bins, each spanning $500$ MC sweeps. The statistical uncertainty is estimated from the standard error obtained using the Jackknife method.
The simulations are performed for system sizes $N=8^3$ and $16^3$ under periodic boundary conditions.

The resulting phases are identified using several observables. First, we compute the spin structure factor defined as
\begin{equation}
    S(\mathbf{q}) = \frac{1}{N}\sum_{l,l'}  \left\langle \mathbf{S}_l \cdot \mathbf{S}_{l'} \right\rangle e^{-i\mathbf{q}\cdot (\mathbf{r}_l - \mathbf{r}_{l'})},
\end{equation}
where $\langle\dots \rangle$ is the average over the MC samples.
Then, the magnetic moments at the ordering vectors $\mathbf{Q}_{\eta}$ are evaluated as
\begin{equation}
    m_{\eta} = \sqrt{\frac{ S(\mathbf{\mathbf{Q}_{\eta}}) }{N}}.
    \label{eq:mq}
\end{equation} 
Similarly, the net magnetization per spin along the magnetic field applied in the $z$ direction is obtained by
\begin{equation}
    m = \frac{1}{N} \left\langle M \right\rangle,
    \label{eq:m}
\end{equation}
with the total magnetization along the $z$ direction $M = \sum_l S^z_l$.
The magnetic susceptibility per spin is evaluated from the fluctuations of $M$ as
\begin{equation}
    \chi_{\mathrm{m}} = \frac{\langle M^2 \rangle - \langle M \rangle^2}{NT}.
    \label{eq:chim}
\end{equation}
Similarly, the specific heat per spin is obtained from the fluctuations of the internal energy as
\begin{equation}
    C =\frac{\langle \mathcal{H}^2 \rangle - \langle \mathcal{H} \rangle^2}{NT^2}.
    \label{eq:C}
\end{equation}

In addition, we evaluate the net scalar spin chirality, which characterizes the noncoplanarity of the spin texture. In itinerant electron systems, the scalar spin chirality can serve as an indicator of emergent magnetic fields relevant to the topological Hall effect~\cite{BINZ20081336,AHE}. It is defined as
\begin{equation}
    \chi_{\mathrm{s}} = \frac{1}{N} \left\langle \sum_{l} \chi_{\mathrm{sc}}^z(\mathbf{r}_l)\right\rangle.
    \label{eq:ssc}
\end{equation}
The local scalar spin chirality $\chi_{\mathrm{sc}}^z$ at $\mathbf{r}_l$ is computed by summing the triple product of neighboring spins over all four triangles surrounding the site in the $xy$ plane as
\begin{equation}
    \chi_{\mathrm{sc}}^z(\mathbf{r}_l) = \frac{1}{2} \sum_{\alpha\beta\nu_\alpha\nu_\beta} \varepsilon^{\alpha\beta} \nu_\alpha\nu_\beta \mathbf{S}_l \cdot (\mathbf{S}_{l+\nu_\alpha\hat{\mathbf{\delta}}_\alpha}\times \mathbf{S}_{l+\nu_\beta\hat{\mathbf{\delta}}_\beta}).
\end{equation}
Here, $\alpha, \beta \in \{x,y\}$, $\varepsilon^{\alpha\beta}$ is the Levi-Civita symbol, $\nu_{\alpha},\nu_\beta=\pm 1$, and $\hat{\mathbf{\delta}}_{\alpha}$ and $\hat{\mathbf{\delta}}_{\beta}$ are the unit translation vectors along the corresponding directions.

To locate magnetic monopoles on the cubic lattice, we employ the discretized solid-angle construction described in Refs.~\cite{yang_hlsimulation, SO_tracingMAM}. We assign a topological charge to each unit cube $C(\mathbf r)$ at lattice position $\mathbf r$ and evaluate the net emergent flux through its boundary $\partial C(\mathbf r)$:
\begin{equation}
    Q_\mathrm{m}(\mathbf{r}) = \frac{1}{4\pi} \sum_{p\subset\partial C(\mathbf r)} \mathbf{\Omega}_p \cdot \hat{\mathbf{n}}_p.
    \end{equation}
The sum extends over the six plaquettes bounding $C(\mathbf r)$, and $\hat{\mathbf n}_p$ is the outward unit normal to plaquette $p$.
The emergent-flux vector associated with each plaquette is defined as
\begin{equation}
    \mathbf{\Omega}_p = \sum_{i \in p} \Omega_i\hat{\mathbf{n}}_p,
    \label{eq:flux}
\end{equation}
where the square plaquette is partitioned into two triangles, and $\Omega_i$ is the oriented solid angle spanned by the three spins $\mathbf S_1$, $\mathbf S_2$, and $\mathbf S_3$ at the vertices of triangle $i$.
The order of the three spins is defined by the outward plaquette normal, which is clockwise when the plaquette is viewed from the center of the cube. The corresponding solid angle is evaluated as
\begin{equation}
    \Omega_i = 2 \operatorname{atan2} \left( \frac{\mathbf{S}_1\cdot (\mathbf{S}_2\times \mathbf{S}_3 )}{1+\mathbf{S}_1\cdot \mathbf{S}_2+\mathbf{S}_2\cdot \mathbf{S}_3+\mathbf{S}_3\cdot \mathbf{S}_1 }\right),
\end{equation} 
Here, $\operatorname{atan2}$ fixes the quadrant of the angle by using the branch for which $\Omega_i\in[-2\pi,2\pi)$. 

With this convention, $Q_\mathrm m=+1$ and $Q_\mathrm m=-1$ identify unit cubes containing a monopole and an antimonopole, respectively. The plaquette fluxes in Eq.~\eqref{eq:flux} can then be used to trace the Dirac string connecting monopoles and antimonopoles, as shown in Fig.~\ref{fig:HL}(a). Each connected monopole–antimonopole pair is identified as a magnetic toron. The total number of torons within a magnetic unit cell $V_\mathrm m$ is therefore calculated as
\begin{equation}
    N_\mathrm{t} = \frac{1}{2}\left \langle \sum_{\mathbf{r}\in V_\mathrm{m}} \left|Q_\mathrm{m}(\mathbf{r})\right|\right\rangle.
    \label{eq:Nt}
\end{equation}

\section{Results}
\label{sec:results}

\begin{figure*}
\includegraphics[width=\textwidth]{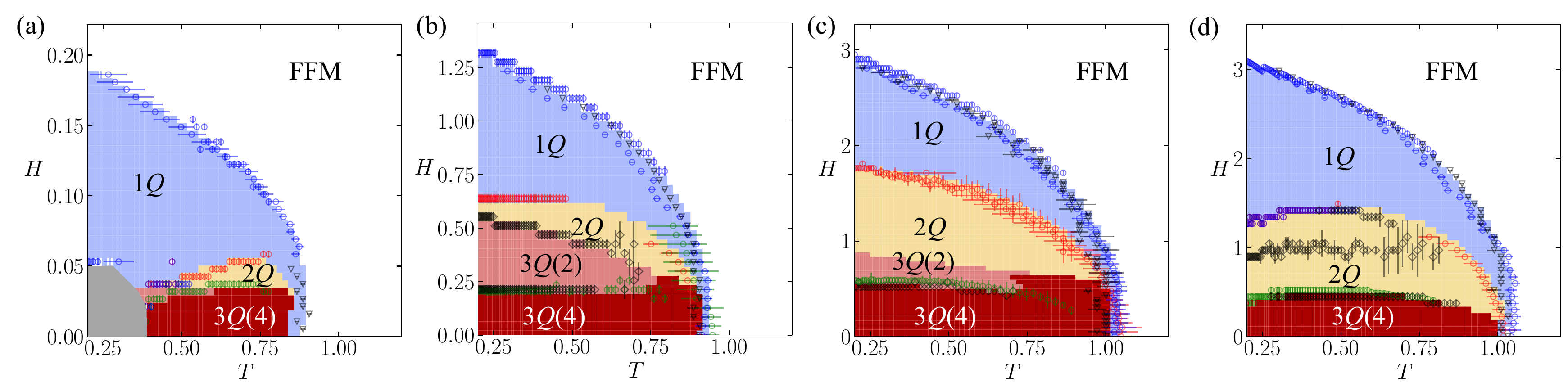}
\caption{\label{fig:HT_pds} $H$-$T$ Phase diagrams for the models with (a) short-range ($\gamma=0.3$ and $r_\mathrm{c}=4$), intermediate-range (b) ($\gamma=0.6$ and $r_\mathrm{c}=6$), and (c) ($\gamma=0.3$ and $r_\mathrm{c}=16$), and (d) infinite-range interactions. The labels $1Q$ and $2Q$ denote the $1Q$ spiral phase and the topologically trivial 2$Q$ phase, respectively; $3Q(4)$ and $3Q(2)$ denote the $3Q$-HL phases, where the numbers in parentheses represent the toron number per magnetic unit cell, and FFM denotes the forced FM phase. In the gray region of panel (a), the phase could not be reliably identified because the MC sampling exhibited pronounced finite-size effects. The phase boundaries are estimated from the peaks in the magnetic susceptibility (diamonds) and specific heat (triangles), as well as from inflection points in the magnetic moments at $\mathbf{Q}_1, \mathbf{Q}_2$, and $ \mathbf{Q}_3$ (red, green, and blue markers, respectively). }
\end{figure*}

We investigate the finite-temperature properties of the models by constructing the field-temperature ($H$-$T$) phase diagrams using parallel-tempering MC simulations.
The main results for $N=16^3$ are summarized in Fig.~\ref{fig:HT_pds}; the resulting phase boundaries are consistent with those obtained for $N=8^3$. 
The phase diagrams are presented in order of increasing interaction range, from the short-range interaction model in Fig.~\ref{fig:HT_pds}(a), through the intermediate-range interaction models in Figs.~\ref{fig:HT_pds}(b) and \ref{fig:HT_pds}(c), to the infinite-range interaction model in Fig.~\ref{fig:HT_pds}(d). Panels (a)-(c) correspond to the finite-range interaction model defined in Eq.~\eqref{eq:H}, while panel (d) corresponds to the infinite-range interaction model defined in Eq.~\eqref{eq:H_infrange}.

The phase boundaries are estimated from the peak positions in the magnetic susceptibility $\chi_\mathrm{m}$ [Eq.~\eqref{eq:chim}] and the specific heat $C$ [Eq.~\eqref{eq:C}], indicated by black diamonds and triangles, respectively. We also use the inflection points of the magnetic moments $m_\eta$ [Eq.~\eqref{eq:mq}] at the ordering wave vectors $\mathbf{Q}_1 = (Q,0,0), \mathbf{Q}_2 = (0,Q,0)$, and $ \mathbf{Q}_3 = (0,0,Q)$, indicated by red, green, and blue markers, respectively. The symbols with horizontal and vertical error bars indicate the phase boundaries determined by the $T$ and $H$ dependences, respectively.
The $3Q$-HL phases are further characterized by the toron number $ N_\mathrm{t}$ [Eq.~\eqref{eq:Nt}].

\subsection{Short-range interaction model}
\label{subsec:short-range}
We first consider the short-range interaction model with $(\gamma,r_\mathrm{c})=(0.3,4)$, whose phase diagram is shown in Fig.~\ref{fig:HT_pds}(a). 
At zero field, the model stabilizes a $3Q$-HL, which contains four torons per magnetic unit cell, quantified as $N_{\mathrm{t}}=4$, and is denoted as $3Q(4)$.
A schematic real-space arrangement of the torons in the $3Q(4)$ phase is shown in Fig.~\ref{fig:HL}(b).
Increasing the temperature drives a continuous phase transition from the $3Q(4)$ phase to the paramagnetic (PM) phase at approximately $T\simeq 0.85$. 
Under an applied magnetic field, the $3Q(4)$ phase undergoes two continuous phase transitions: first to the topologically trivial $2Q$ phase and then to the $1Q$ phase.
Reentrant behavior is observed at the phase boundaries between the $2Q$ and $1Q$ phases for $0.03 \lesssim H \lesssim 0.05$.
The $1Q$ phase extends over a comparatively wide region of the phase diagram and continuously evolves into the forced FM (FFM) phase. The FFM phase is a PM phase whose symmetry is broken by the external magnetic field, and is continuously connected to the PM phase at zero field.
The low-temperature and low-field region could not be reliably assigned, because the MC sampling exhibited pronounced finite-size effects; this region is therefore shaded in gray in Fig.~\ref{fig:HT_pds}(a). 

We compare our results to those obtained for the short-range interaction model studied in Ref.~\cite{yang_hlsimulation}, which includes exchange and DM interaction up to third-nearest neighbors. It was shown that a $4Q$-HL can be stabilized at a finite magnetic field depending on the modulation wavelength.
By contrast, the short-range interaction model considered here stabilizes a $3Q$-HL even at zero field. This discrepancy indicates that the formation of HLs in the short-range case depends sensitively on the specific interaction profile. 

\subsection{Infinite-range interaction model}
\label{subsec:long-range}
Next, we consider the opposite limit, namely the infinite-range interaction model, whose phase diagram is shown in Fig.~\ref{fig:HT_pds}(d). The $3Q$-HL phase is identified for magnetic fields up to $H\simeq 0.4$. 
At low temperatures, increasing the magnetic field drives successive discontinuous transitions from the $3Q$-HL phase to the $2Q$ phase and then to the $1Q$ phase.
At higher fields, the $1Q$ phase continuously evolves into the FFM phase. 
By increasing the temperature, the discontinuities at the phase transitions among the $3Q$-HL, $2Q$, and $1Q$ phases gradually become weaker, eventually becoming continuous above $T\simeq 0.5$. The $3Q$-HL and $1Q$ phases undergo continuous phase transitions into the PM and FFM phases, respectively.
Furthermore, reentrant behavior is observed by lowering the temperature from $T_\mathrm c \simeq 1.0$ towards $0.2$ at the boundaries between the $3Q$-HL and $2Q$ phase, as well as between the  $2Q$ and $1Q$ phases. 

For the infinite-range interaction model considered here, a steepest-descent analysis was previously carried out in Ref.~\cite{PhysRevB.105.174413}. We confirm that our MC results are consistent with those obtained from that analysis (not shown), both in terms of the phase diagram and the reentrant behavior between the $3Q$-HL and $2Q$ phases, as well as between the $2Q$ and $1Q$ phases.
A distinctive feature of the infinite-range interaction case, not reported in Ref.~\cite{PhysRevB.105.174413}, is an anomaly in the magnetic susceptibility within the $2Q$ phase at finite temperatures below $T \simeq 0.75$.  
This anomaly coincides with the region where the toron number shows nonzero values within the otherwise topologically trivial $2Q$ phase. The full $H$ and $T$ dependence of the toron number is shown in Fig.~\ref{fig:Nt_infrange} of Appendix~\ref{app:disordered_HL}.
This anomaly indicates the presence of a regime at $H \lesssim 1.0$, in which torons are thermally generated under a magnetic field. Since these torons are created and annihilated by thermal fluctuations and do not form a periodic lattice, this regime may be interpreted as a disordered magnetic toron state. 
This difference may originate from fluctuation-driven, low-energy collective excitations, which can be captured by MC simulations through direct configuration sampling, but cannot be directly resolved within the mean-field framework.

\subsection{Intermediate-range interaction models}
\label{subsec:intermediate-range}

\begin{figure}
\includegraphics[width=\columnwidth]{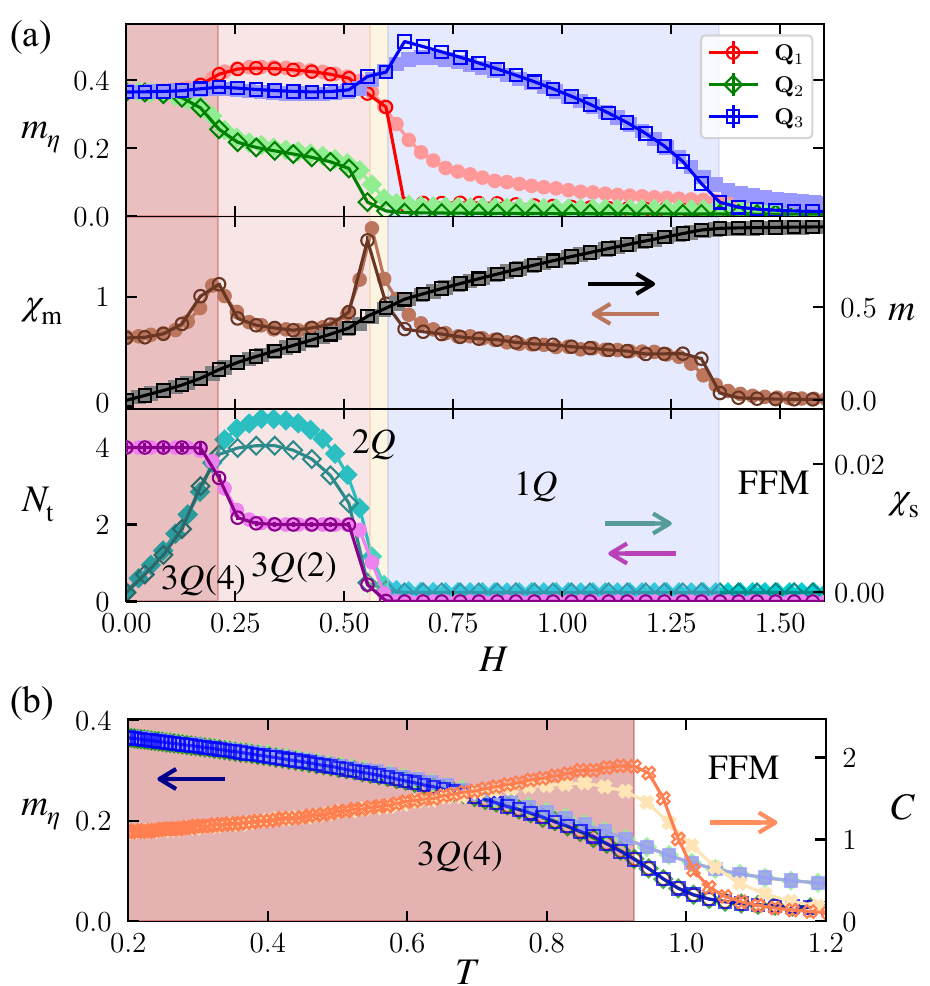}
\caption{\label{fig:App_HT_scan}(a) $H$ dependences at $T=0.2$ of (top) the magnetic moments $m_{\eta}$ at the ordering wave vectors $\mathbf{Q}_1, \mathbf{Q}_2$, and $\mathbf{Q}_3$, indicated by red, green, and blue markers, respectively; (middle) the magnetic susceptibility $\chi_{\mathrm{m}}$ and net magnetization per spin $m$; and (bottom) the toron number $N_{\mathrm{t}}$ and scalar spin chirality $\chi_{\mathrm{s}}$ for the intermediate-range interaction model with $\gamma = 0.6$ and $r_{\mathrm{c}} = 6$.
(b) $T$ dependences at $H=0$ of $m_{\eta}$ and the specific heat $C$ for the same model. Filled markers with pale colors and open markers with dark colors represent the results obtained for the system sizes $N = 8^3$ and $16^3$, respectively.}
\end{figure}
\begin{figure*}
\includegraphics[width=\textwidth]{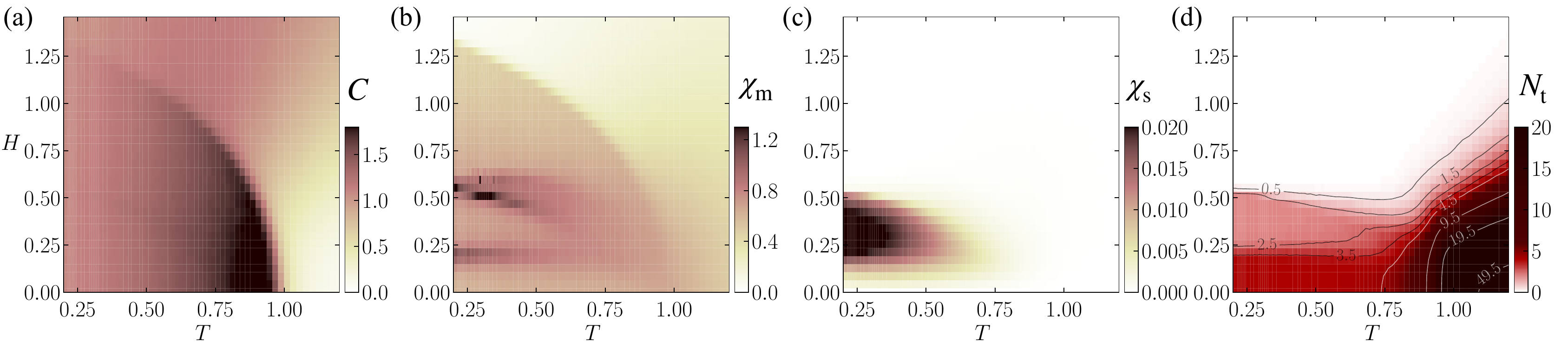}
\caption{\label{fig:HT_observables} $H$ and $T$ dependences of (a) the specific heat, (b) the magnetic susceptibility, (c) the scalar spin chirality, and (d) the toron number for the intermediate-range interaction model with $\gamma=0.6$ and $r_\mathrm{c}=6$.}
\end{figure*}
 
Finally, we turn to the central result of this study, namely the phase diagrams for intermediate-range interaction models. Figures~\ref{fig:HT_pds}(b) and \ref{fig:HT_pds}(c) show the results for $(\gamma,r_\mathrm{c})=(0.6,6)$ and $(0.3,16)$, respectively. 
A qualitative difference from both the short- and infinite-range interaction models is that a second $3Q$-HL phase, $3Q(2)$, is stabilized upon applying a magnetic field \cite{shimizu2025currentinduced-01a}.

We first focus on Fig.~\ref{fig:HT_pds}(b). 
At zero magnetic field, the stable phase is the $3Q(4)$ state. Upon applying a magnetic field, the system undergoes a transition to the topologically distinct $3Q(2)$ phase. 
This transition can be understood as follows. Starting from the $3Q(4)$ state, the magnetic field drives the hedgehogs and antihedgehogs toward one another by shortening the skyrmion strings connecting them.
Since two of the four torons have shorter skyrmion strings [Fig.~\ref{fig:HL}(b)], they pair annihilate at $H\simeq 0.2$, resulting in a $3Q$-HL phase with $N_{\mathrm{t}} = 2$, denoted as $3Q(2)$.
Upon further increasing the field, the system undergoes successive transitions to the $2Q$ phase and then to the $1Q$ phase. These transitions appear to be discontinuous at low temperatures (see below), whereas they become continuous at higher temperatures.
The transitions from the $1Q$ to the FFM phase and from the $3Q$-HL to the PM phase at zero field are continuous, as in the infinite-range interaction model.

Figure~\ref{fig:HT_pds}(c) shows the phase diagram for a longer but still finite interaction range. The set of stabilized phases and the overall sequence of phase transitions are qualitatively similar to those in Fig.~\ref{fig:HT_pds}(b). 
The principal difference lies in the relative stability of the phases: the $3Q(2)$ phase is narrowed, whereas the topologically trivial $2Q$ phase occupies a larger region of the phase diagram.

In the following, we examine the observables obtained from the MC sampling to provide a clearer picture of the phase structures.
Figure \ref{fig:App_HT_scan} shows the field scan at $T=0.2$ and the temperature scan at zero field, corresponding to the phase diagram in Fig.~\ref{fig:HT_pds}(b).
The open and filled markers denote the results obtained for the system sizes $N=16^3$ and $N=8^3$, respectively. The corresponding results for Fig.~\ref{fig:HT_pds}(c) are discussed in Appendix~\ref{app:ht_intermedrange2}.

The upper panel of Fig.~\ref{fig:App_HT_scan}(a) shows the three magnetic moments $m_\eta$ at the ordering wave vectors $\mathbf{Q}_1, \mathbf{Q}_2$, and $ \mathbf{Q}_3$. 
At zero field, the $3Q(4)$ phase preserves $C_3$ symmetry about the $[111]$ axis, as reflected in the equal intensities of the magnetic moments.
The same behavior is observed in the temperature scan in Fig.~\ref{fig:App_HT_scan}(b). 
The specific heat, also shown in the same panel, exhibits an anomaly at $T\simeq 0.9$, suggesting a continuous phase transition to the PM phase.

An applied magnetic field breaks the $C_3$ symmetry, leading to unequal magnetic moments with the hierarchy $m_1 > m_3 > m_2 \geq 0$, as shown in the upper panel of Fig.~\ref{fig:App_HT_scan}(a).
At $H \simeq 0.2$, the system undergoes a transition into the $3Q(2)$ phase. The magnetic susceptibility $\chi_{\mathrm{m}}$, shown in the middle panel, exhibits a broad hump-like anomaly around the same field, while the net magnetization per spin $m$ [Eq.~\eqref{eq:m}] shows a weak anomaly. 

The distinction between the two $3Q$-HL phases is captured by the toron number $N_\mathrm{t}$ per magnetic unit cell, shown in the lower panel of Fig.~\ref{fig:App_HT_scan}(a). 
At low temperatures, $N_\mathrm{t}$ changes rapidly from nearly $4$ to $2$ at $H\simeq 0.2$, which indicates that the transition is accompanied by a change in topological nature. 
The lower panel also shows the scalar spin chirality $\chi_{\mathrm{s}}$ [Eq.~\eqref{eq:ssc}]. At zero magnetic field, $\chi_{\mathrm{s}}$ vanishes owing to the high symmetry of the $3Q(4)$ state. When a magnetic field is applied, $\chi_{\mathrm{s}}$ becomes finite in the $3Q(4)$ phase, increases rapidly near the topological transition, and reaches its maximum in the $3Q(2)$ phase. It then gradually decreases and eventually vanishes upon entering the $2Q$ phase. Thus, at finite field, $\chi_{\mathrm{s}}$ serves as a measure of the topological nature of the $3Q$-HLs \cite{AHE}.
These features suggest that the transition from the $3Q(4)$ phase to the $3Q(2)$ phase involves changes in both the topological character and magnetic order.

At the transition from the $3Q(2)$ phase to the $2Q$ phase, $m_2$ jumps discontinuously to zero, which becomes more pronounced with increasing system size. Additionally, $\chi_{\mathrm{m}}$ exhibits a pronounced peak and $m$ shows a small jump, suggesting a first-order phase transition. At the subsequent transition from the $2Q$ phase to the $1Q$ phase, $m_1$ likewise vanishes discontinuously, leaving only $m_3$ finite, also suggesting a first-order transition. At higher fields, $m_3$ continuously decreases to zero as the system enters the field-polarized FFM regime, accompanied by an anomaly in $\chi_{\mathrm{m}}$. This behavior is consistent with a continuous transition.

The full $H$-$T$ dependences of the observables are presented in Fig.~\ref{fig:HT_observables}.
Figure \ref{fig:HT_observables}(a) shows the specific heat, which exhibits pronounced enhancements at the phase boundaries between the $1Q$ and the FFM phase and between the $3Q(4)$ and the PM phase. 
The magnetic susceptibility in Fig.~\ref{fig:HT_observables}(b) displays clear anomalies at the boundaries between the two $3Q$-HL phases, between the $3Q(2)$ and $2Q$ phases, as well as between the $2Q$ and $1Q$ phases.
Figure \ref{fig:HT_observables}(c) shows the scalar spin chirality, which takes nonzero values only in the $3Q$-HL phases. 

Figure \ref{fig:HT_observables}(d) shows the toron number $N_{\mathrm{t}}$, whose magnitude is visualized by both the color scale and contour map.
$N_\mathrm{t}$ takes values of nearly $4$ in the $3Q(4)$ phase and $2$ in the $3Q(2)$ phase.  
Upon approaching the critical temperature $T_\mathrm c\simeq 0.9$, $N_{\mathrm{t}}$ increases owing to thermal fluctuations and can reach values corresponding to a substantial fraction of the system size in the PM phase. In this regime, hedgehogs and antihedgehogs are thermally proliferated without forming a periodic structure. For visibility, the color scale of $N_{\mathrm{t}}$ in Fig.~\ref{fig:HT_observables}(d) is restricted to $N_{\mathrm{t}}\leq 20$. As the magnetic field increases, $N_\mathrm{t}$ is gradually suppressed and eventually vanishes in the field-polarized region.
\section{Discussion}
\label{sec:discussion}

Based on these results, the phase diagrams can be classified into three qualitatively distinct regimes. In the short-range interaction regime, only the $3Q(4)$ phase is stabilized among the $3Q$-HL phases. In the intermediate-range regime, two topologically distinct $3Q$-HL phases, $3Q(4)$ and $3Q(2)$, appear over finite regions of the $H$-$T$ phase diagram. In the infinite-range regime, the $3Q(2)$ phase is destabilized, whereas the topologically trivial $2Q$ phase becomes more prominent. 
Thus, the stability of the HL phases does not vary monotonically with the interaction range. Among the interaction profiles examined here, only the intermediate-range interaction models stabilize both the $3Q(4)$- and $3Q(2)$-HL phases.
These results suggest that a finite, but sufficiently extended interaction range appears to be required to stabilize the field-induced $3Q(2)$ phase, whereas further increasing the interaction range favors the topologically trivial $2Q$ phase.

The absence of the $3Q(2)$ phase in the short- and infinite-range limits may be understood from the momentum-space interaction profile. 
Decreasing the interaction range broadens the interaction profile around the ordering wave vectors $\mathbf{Q}_\eta$ in momentum space, as shown in Fig.~\ref{fig:Jr_Jq}(b), thereby allowing a wider range of $\mathbf q$ components to contribute to the spin texture. 
At the same time, however, the interactions for $(\gamma,r_{\mathrm{c}})=(0.3,4)$ retain a substantial weight at $\mathbf q=0$, which is associated with FM order. 
Since an external magnetic field directly enhances this uniform component, the short-range interaction model tends to favor field-induced spin polarization, thereby destabilizing the HL phases. This tendency may explain why the $3Q(4)$ phase transforms directly into the topologically trivial $2Q$ phase, without an intervening $3Q(2)$ phase.

By contrast, the momentum-space interaction profile of the infinite-range interaction model contains contributions only from the ordering vectors. 
This restriction may make the spin texture more rigid, preventing magnetic hedgehogs from flexibly rearranging their positions under an external magnetic field. As a result, the system may favor a $2Q$ structure with dominant $Q$ vector parallel to the external field, rather than stabilizing the $3Q(2)$ phase. 

The intermediate-range interaction models lie between these two limits. Compared to the short-range interaction model, they have reduced interaction weight near $\mathbf{q}=0$, while retaining a finite width around the ordering wave vectors. This balance may provide sufficient flexibility for the spin texture and monopole–antimonopole configuration to deform under the magnetic field without excessively favoring uniform polarization, thereby stabilizing the field-induced $3Q(2)$ phase.
A related explanation concerns the shape of the interaction in momentum-space itself: in some systems, $q^4$ terms in the interaction can emerge within a purely bilinear (two-spin) exchange model once the real-space interaction decays with distance~\cite{rybakovhopfions2022,PhysRevLett.108.017206}. Because biquadratic (four-spin) interactions are known to stabilize multiple-$Q$ states~\cite{Hayami_2021, PhysRevLett.108.096401, PhysRevB.95.224424, PhysRevB.101.144416, 10.7566/JPSJ.91.093702}, such decay-induced $q^4$ terms may play an analogous role in stabilizing the different $3Q$-HLs discussed here.

\section{Conclusion}
\label{sec:conclusion}

In summary, we systematically investigated how the spatial range of bilinear exchange interactions affects the stability of magnetic HLs. Our MC simulations show that the $3Q(4)$-HL remains stable over a broad range of interaction profiles, from short-range to infinite-range limits. Among the profiles examined, intermediate-range interactions are particularly favorable for stabilization of HLs, because they additionally stabilize a field-induced $3Q(2)$ phase. The relative stability of the $3Q(4)$, $3Q(2)$, and topologically trivial $2Q$ phases is therefore highly sensitive to the real-space decay of the effective interactions or, equivalently, to their peak width in momentum space.

Our framework bridges interaction profiles commonly used for localized-spin insulating models and itinerant-electron metallic models, and provides a basis for interpreting topological magnetic phase diagrams in candidate materials. In itinerant magnets, doping, gating, and pressure may modify carrier-mediated exchange through changes in the electronic structure and spin susceptibility \cite{PhysRevLett.91.087205, PhysRevLett.87.217201, Chongthanaphisut2022, Sokolov2009}. In localized-spin magnets, pressure, chemical substitution, and interface engineering may alter competing superexchange pathways \cite{PhysRevB.92.054413, PhysRevLett.124.077202, 10.1038/s41563-019-0506-1, doi:10.1126/science.aav1937}.
Furthermore, a continuum microscopic model could be constructed, especially for the intermediate-range interaction models~\cite{S_J_M_J_O_G_S_Y_S_2020, PhysRevLett.128.157206}, which could give further insights on the stability criteria for HL states, enable dynamical studies based on the Landau–Lifshitz–Gilbert equation, and facilitate calculations of excitation spectra.  
In parallel, the model interactions could be benchmarked against density functional theory calculations to assess whether the intermediate-range interaction model is realistic for candidate materials. 
Further work should also investigate higher-order spin interactions, which may stabilize $4Q$-HL phases.

\begin{acknowledgments}
We thank K. Kobayashi, K. Okigami, and R. Yambe for fruitful discussions. This work was supported by the JSPS KAKENHI (Grants No. JP22K13998, No. JP23K25816, No. JP25H01247, No. JP26K07011, and No. JP26H00634) and JST PRESTO (Grant No. JPMJPR2595). M.Y. was supported by the Forefront Physics and Mathematics program to drive transformation (FoPM). The computation in this work has been done using the facilities of the Supercomputer Center, the Institute for Solid State Physics, The University of Tokyo.
\end{acknowledgments}

\appendix

\section{Details of finite-range interactions} 
\label{app:interaction}

\begin{figure}
\includegraphics[width=\columnwidth, keepaspectratio]{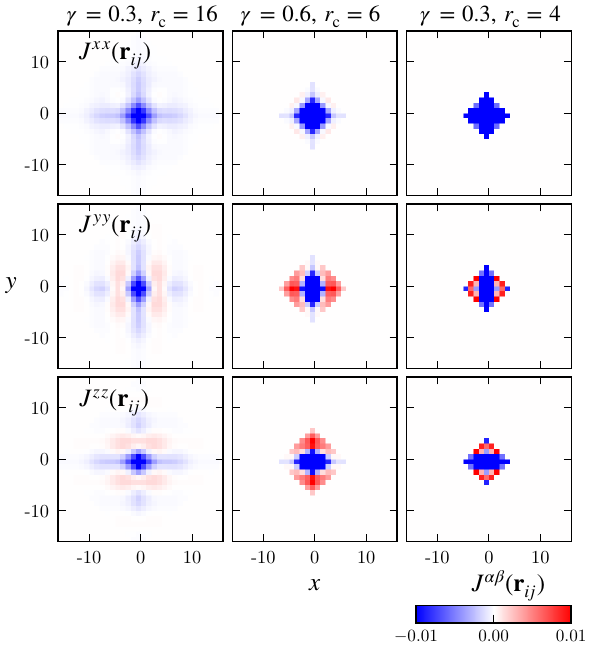}
\caption{\label{fig:Jx_all} Color maps of the exchange interaction $J^{\alpha\alpha}(\mathbf{r}_{ij})$ for $\alpha = x,y,z$ evaluated on the $z=0$ plane, with $\mathbf{r}_{ij} = (x, y, 0)$, for the intermediate-range ($\gamma=0.3$ and $r_{\mathrm{c}} =16$) (left column), and ($\gamma=0.6$ and $r_\mathrm{c} =6$) (center column) and short-range ($\gamma=0.3$ and $r_{\mathrm{c}} =4$) (right column).}
\end{figure}

\begin{figure}
\includegraphics[width=\columnwidth, keepaspectratio]{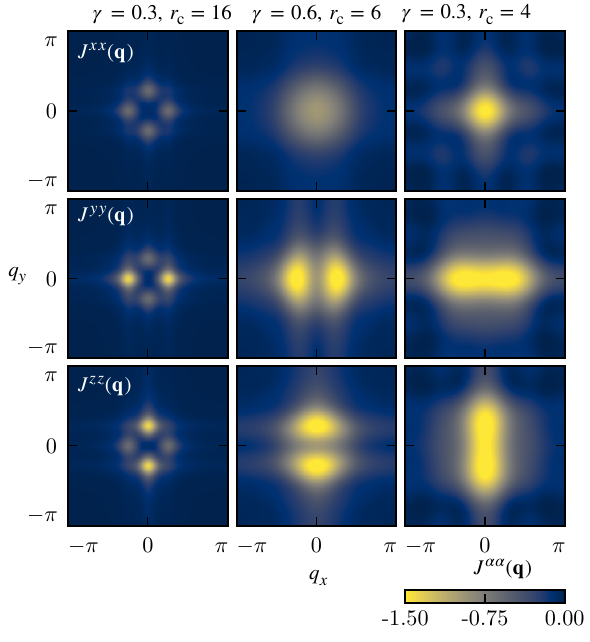}
\caption{\label{fig:Jq_all} Color maps of the exchange interactions $J^{\alpha\alpha}(\mathbf{q})$ for $\alpha=x,y,z$ on the $q_z=0$ plane, with $\mathbf{q}=(q_x,q_y,0)$, for the intermediate-range ($\gamma=0.3$ and $r_\mathrm{c} =16$) (left column), and ($\gamma=0.6$ and $r_{\mathrm{c}} =6$)  (center column) and short-range ($\gamma=0.3$ and $r_{\mathrm{c}} =4$) (right column).}
\end{figure}

Figure~\ref{fig:Jx_all} shows real-space color maps of the symmetric exchange interactions $J^{\alpha\alpha}(\mathbf{r}_{ij})$ for $\alpha = x,y,z$ [Eq.~\eqref{eq:Jij_finite}]. The interactions are evaluated on the $z=0$ plane, with $\mathbf{r}_{ij} = (x, y, 0)$. 
Results are shown for three sets of parameters used in the present study: $(\gamma,r_\mathrm{c})= (0.3, 16), (0.6, 6)$, and $(0.3,4)$.
As discussed in Sec.~\ref{subsec:model} and visually evident in Fig.~\ref{fig:Jx_all}, the effective interaction range defined in Eq.~\eqref{eq:reff} decreases in this order. 
Reducing the cutoff length $r_{\mathrm c}$ truncates the interactions more abruptly, producing the characteristic diamond-shaped patterns due to the distance measured by the Manhattan distance $g(\mathbf{r}_{ij})= |r_{ij}^x| + |r_{ij}^y| + |r_{ij}^z|$. Line cuts along the $x$ axis at $y=z=0$ are presented in Fig.~\ref{fig:Jr_Jq}(a).
Although the left and right columns have the same decay parameter $\gamma=0.3$, their interaction profiles differ even within the cutoff $r_\mathrm{c}$ because $Q^*$ is optimized separately for each value of $r_\mathrm{c}$.

Figure~\ref{fig:Jq_all} presents the corresponding momentum-space color maps of $J^{\alpha\alpha}(\mathbf{q})$ for the same sets of interaction-range parameters on the $q_z=0$ plane. These interactions are obtained by Fourier transforming Eq.~\eqref{eq:Jij_finite}.
As the interaction range increases, the peaks centered at the ordering wave vectors become sharper. 
The $q_x$ line cuts at $q_y=q_z=0$ correspond to those shown in Fig.~\ref{fig:Jr_Jq}(b).

\section{Toron number in the infinite-range interaction model}
\label{app:disordered_HL}
\begin{figure}
\includegraphics[width=0.9\columnwidth]{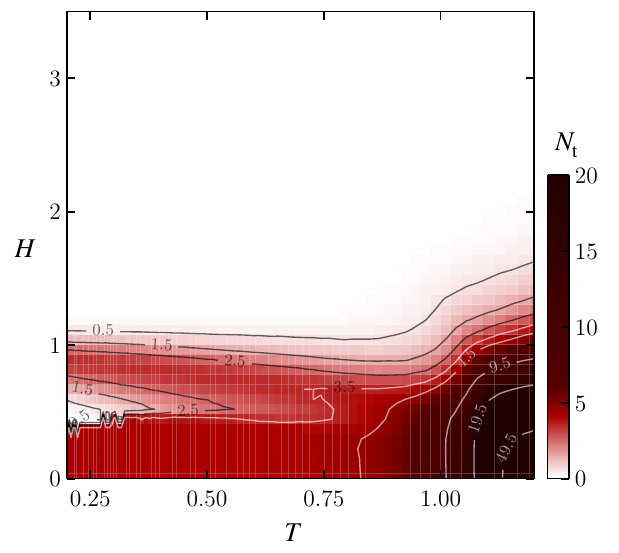}
\caption{\label{fig:Nt_infrange} $H$-$T$ dependence of the toron number $N_{\mathrm{t}}$ for the infinite-range interaction model.
}
\end{figure}
In this appendix, we present the $H$-$T$ dependence of the toron number $N_\mathrm{t}$ per magnetic unit cell in the infinite-range interaction model, visualized in Fig.~\ref{fig:Nt_infrange}.
The color scale indicates $N_\mathrm{t}$, which is $\simeq 4$ in the $3Q(4)$ phase. 
A region with finite $N_\mathrm{t}$ appears for $0.7 \lesssim H \lesssim 1.0$ at low temperatures and extends up to the critical temperature. This region coincides with the anomaly in the magnetic susceptibility plotted by diamonds in Fig.~\ref{fig:HT_pds}(d).
Remarkably, this region lies within the topologically trivial $2Q$ phase, characterized by the finite magnetic moments $m_2$ and $m_3$. 
In this regime, the finite toron number is associated with magnetic torons that are created and annihilated by thermal fluctuations, but do not form a periodic lattice. This regime may be interpreted as a disordered magnetic toron state. Among the interaction profiles examined, such a regime appears only in the infinite-range interaction model.

\section{$H$ and $T$ scan of the intermediate-range interaction model with $\gamma=0.3$ and $r_\mathrm{c}=16$}
\label{app:ht_intermedrange2}

\begin{figure}
\includegraphics[width=\columnwidth]{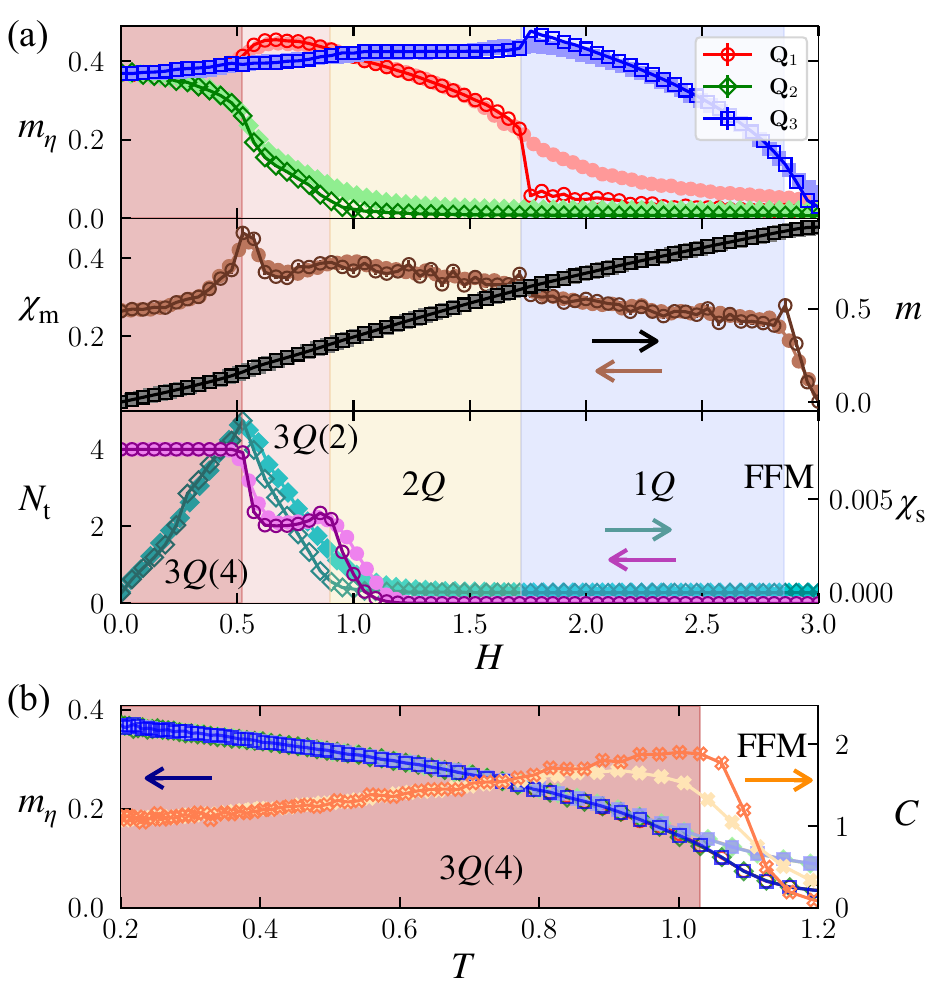}
\caption{\label{fig:htscan_g03rc16}(a) $H$ dependences at $T=0.2$ of (top) the magnetic moments $m_{\eta}$ at the ordering wave vectors $\mathbf{Q}_1, \mathbf{Q}_2$, and $\mathbf{Q}_3$, indicated by red, green, and blue markers, respectively; (middle) the magnetic susceptibility $\chi_{\mathrm{m}}$ and net magnetization $m$; and (bottom) the toron number $N_{\mathrm{t}}$ and scalar spin chirality $\chi_{\mathrm{s}}$ for the intermediate-range interaction model with $\gamma = 0.3$ and $r_{\mathrm{c}} = 16$.
(b) $T$ dependences at $H=0$ of $m_{\eta}$ and the specific heat $C$ for the same model.
The notations are common to those in Fig.~\ref{fig:App_HT_scan}.}
\end{figure}

Figure~\ref{fig:htscan_g03rc16} shows the field scan at $T=0.2$ and the zero-field temperature scan for the intermediate-range parameter set $(\gamma,r_\mathrm{c}) =(0.3, 16)$. These scans correspond to the phase diagram in Fig.~\ref{fig:HT_pds}(c) and use the same panel layout and plotting conventions as Fig.~\ref{fig:App_HT_scan}. 
Compared with the results in Fig.~\ref{fig:App_HT_scan}, the $3Q(2)$ phase is reduced, whereas the $2Q$ phase is enlarged. This suggests that longer-range interactions may make the spin texture more rigid under an applied magnetic field, preventing magnetic hedgehogs from flexibly rearranging their positions and instead favoring the $2Q$ phase. 
Similar discussions in a broader context are provided in Sec.~\ref{sec:discussion}.

\newpage

\bibliography{aapmsamp}

\end{document}